\documentclass[aps,nofootinbib,superscriptaddress,twocolumn]{revtex4}

\usepackage{amsmath}
\usepackage{graphicx}
\usepackage[figuresright]{rotating}

\begin{document}
\title{Inclusive nucleon emission 
induced by quasi-elastic neutrino-nucleus interactions}

\author{J. Nieves}
\affiliation{Departamento de F\'\i sica Moderna, \\
Universidad de Granada, 
Campus de Fuentenueva, S/N, E-18071, Granada, Spain},
\author{M. Valverde}
\affiliation{Departamento de F\'{\i}sica Moderna,\\
  Universidad de Granada,
  Campus de Fuentenueva, S/N,
  E-18071 Granada, Spain}
\author{M.J. Vicente-Vacas}
\affiliation{Departamento de F\'{\i}sica Te\'{o}rica and IFIC\\
  Centro Mixto Universidad de Valencia-CSIC,
  46100 Burjassot (Valencia), Spain}
       
\begin{abstract}
A previous model on inclusive charged-current quasi-elastic nuclear reactions 
\cite{Nie04} is extended to include neutral- and charged-current nucleon
emission reactions. The problem of outgoing nucleon propagation is treated 
by means of a Monte Carlo simulation.
\vspace{1pc}
\end{abstract}

\maketitle

\section{INTRODUCTION}

The main effort of theoretical studies on neutrino-nucleus reactions
at intermediate energies (few hundreds MeV) has been 
devoted to quasi-elastic (QE) charged-current (CC) reactions. This is due to 
the fact that water Cerenkov detectors only detect outgoing charged leptons.
New experiments such as K2K aim to measure the espectrum of outgoing nucleons
produced in neutral currents (NC) in this energy range. These kind of 
observables are also of interest when dealing with the extraction of 
information on
strange quark axial content of the nucleon. For a correct theoretical
anlysis of these experiments we have to take into account the rescattering 
effect of the outgoing nucleons in its way out of the nucleus. 
 
In \cite{Nie04} a model dealing with QE CC reactions of the
type $^A_ZX(\nu_l,l^-)Y$ was presented. This model was based in a local Fermi
gas model of the nucleus and included further nuclear effects in a parameter
free fashion, thus leading to one of the best descriptions of 
the LSND \cite{LSND} experiments.
Our confidence in this model at intermediate energies higher than those of 
\cite{LSND} is based on the fact that a
similar model \cite{GNO97} is the only model on electron scattering that 
has been successfully compared with experiment from the QE to the
$\Delta$ excitation regions.

In section \ref{SeccNC} we extend the model in \cite{Nie04} to NC 
$^A_ZX(\nu_l,\nu_l)Y$. Section \ref{SeccMC} deals with the Monte Carlo 
treatment of the rescattering problem. Finally we show results for outgoing 
nucleon observables in CC and NC.

\section{NEUTRAL CURRENTS} \label{SeccNC}

In NC processes the $Z^0$ boson can be absorbed by one
nucleon leading to the QE contribution of the nuclear response function.
\begin{equation}
\nu_l(k) + ^A_ZX \to \nu_l (k^\prime)+ Y
\label{eq:reac}
\end{equation}
We obtain the cross section of this proccess by calculating the imaginary part
of the self-energy diagram associated to the $Z^0$ absorption in nuclear 
matter. Results for finite nuclei are obtained with the local density 
approximation. This method was also used in \cite{Nie04} with CC processes, 
the main differences in NC arising from the vertex $Z^0NN$
\begin{equation}
\left\langle N; {\bf p}^{\prime}={\bf p}+{\bf q}\left| j^\alpha_{\text{nc}}(0) 
\right| N;{\bf p}\right\rangle =
\bar{u}({\bf p}{^\prime})(V^\alpha_N-A^\alpha_N)u({\bf p}) 
\end{equation}
with vector and axial nucleon NC given by
\begin{gather}
V^\alpha_N = 2 \times  \left ( F_1^Z(q^2)\gamma^\alpha + i 
\mu_Z \frac{F_2^Z(q^2)}{2M}\sigma^{\alpha\nu}q_\nu\right)_N, \\
A^\alpha_N = \left ( G_A^Z(q^2) \gamma^\alpha\gamma_5   + G_P^Z(q^2)
 q^\alpha\gamma_5 \right)_N \, .
\end{gather}
Thanks to SU(3) symmetry some relations 
exist among the NC form factors and the CC ($F_{1,2}^V$, $G_A$ and
$G_P$) and the electromagnetic ones. We have to 
introduce $F_1^s, \mu_s F_2^s, G_A^s$ and $G_P^s$, the strange vector
and axial nucleon form factors \cite{Al96}. We use the parametrizations
of \cite{Ga71} for electromagnetic form factors and those of fit II 
in \cite{Ga93} for strange ones.

Like in \cite{Nie04} further nuclear effects were taken into account, mainly 
keeping a correct energy balance and including long range correlations 
(RPA). The main differences with respect to CC arise 
from the inclusion of isoscalar pieces in the nucleon-nucleon effective 
interaction and the absence of the outgoing lepton Coulomb interaction
with the final nucleus.

\section{MONTE CARLO SIMULATION}\label{SeccMC}

In the previous section we overviewed the evaluation of NC inclusive neutrino
induced cross sections in the QE region. Thus we determined, for a fix incoming
neutrino energy, the inclusive QE cross section $d^2\sigma /d\Omega^{\prime}
dE^{\prime}$ ($\Omega^{\prime}$, $E^{\prime}$ solid angle and energy
of the outgoing lepton). In our scheme this is obtained after performing an 
integration over the whole nuclear volume, following the line of the local 
density approximation of the Fermi gas model. Thus, for a fix transferred
four momentum $q^\mu$, chosen according to $d^2\sigma
/d\Omega^{\prime} dE^{\prime}$, we can randomly select the point of
the nucleus where the absorption takes place using the profile
$d^5\sigma /d\Omega^{\prime} dE^{\prime}d^3{\bf r}$.

Now, we need to have on top of that the distribution of
three--momenta of the outgoing nucleon. Since events are generated 
probabilistically one by one, we generate a random momentum from the local
Fermi sea, and this will generate an outgoing momentum $p^{\prime}=
p+q$. If it happens that $\left|{\bf p}^{\prime}\right|<k_{F}(r)$ (the local
Fermi momentum) then the event is Pauli blocked, it is dismissed and
another event is generated. Thus, we have already the configuration of
the final state after the first step: in this case just one nucleon
produced in the point {\bf r} of the nucleus with momentum
${\bf p}^{\prime}$.  As with respect to having a proton and a
neutron in the final state, this is trivially done: for the CC case,
the outgoing nucleon is a proton (neutron) for neutrino (antineutrino)
induced processes, while for NC the reaction
probability was already splitted into a proton and a neutron induced
ones.
 
Then we simulate the trajectory of the ejected nucleon in its way out of the
nucleus. We move the nucleon in finite length steps ($d << \lambda$, 
the nucleon mean free path). At every new position we evaluate the probability
of the nucleon colliding so it produces a new outgoing secondary nucleon. 
All of the ejected nucleons (the primary one, and the possibly existing 
secondary) are again propagated until they all leave the nucleus, all of them
giving the same contribution to the differential cross section.

When a QE nucleon-nucleon collision is produced a random Fermi sea nucleon
momentum is selected. The new primary and secondary nucleon kinematics are
selected according to experimental differential cross sections of free space
nucleon-nucleon collision, conveniently modified in order to take into account
the effects of medium polarization (correlations) and Pauli blocking effects.
The propagation of nucleons is simulated with a semiclassical approximation
of nucleons moving in a nucleon-nucleus potential. 
This approximation was shown to be fairly accurate for nucleons with kinetic 
energies higher than $\simeq 20$ MeV in \cite{COV94}.

\section{RESULTS}

\begin{figure}[htb]\begin{center}
\includegraphics[scale=0.3]{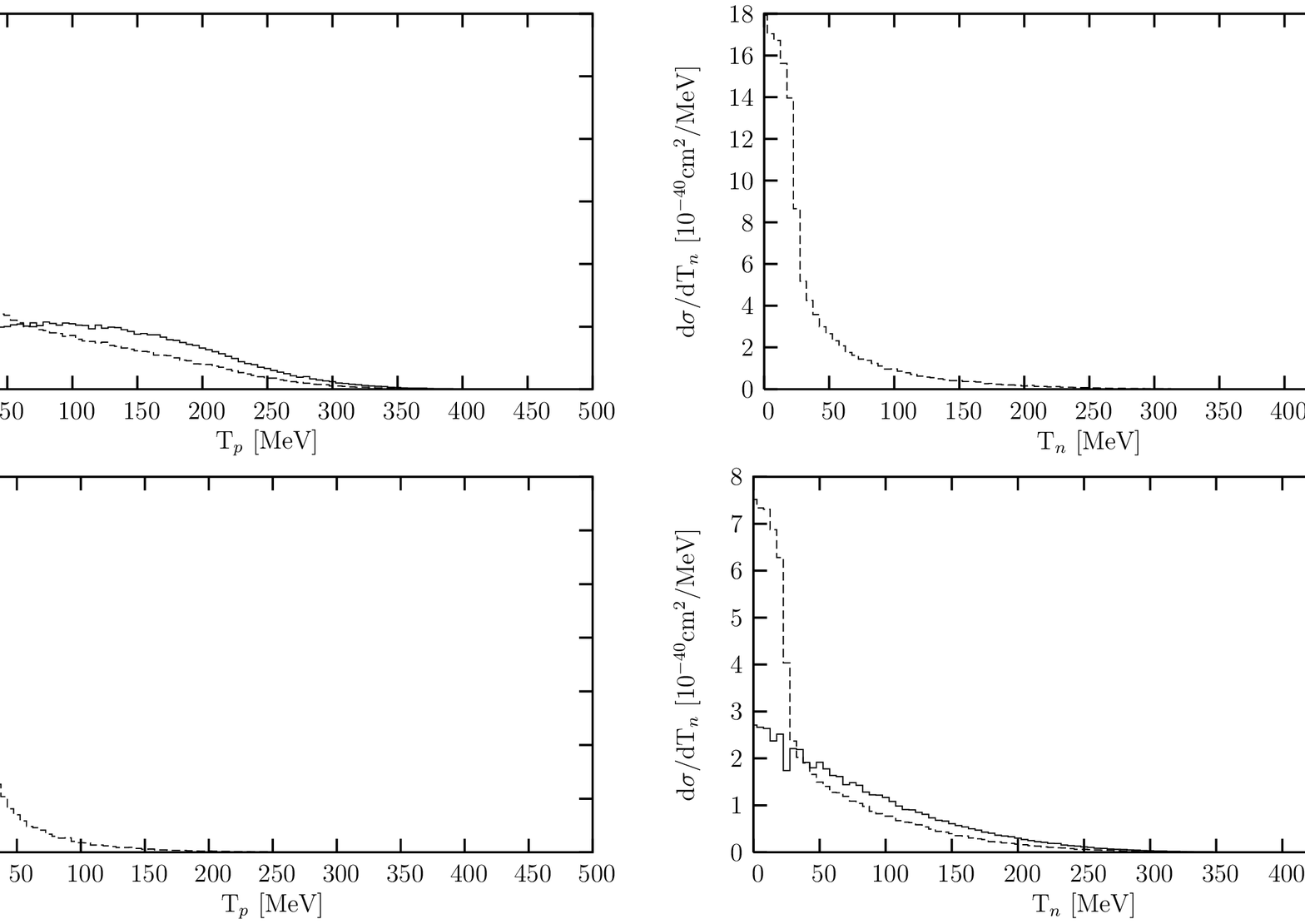}
\caption{\footnotesize 
Charged current $^{40}Ar(\nu,\mu^-+N)$ (upper panels) and 
$^{40}Ar(\bar{\nu},\mu^++N)$ (lower panels) cross sections as a function
of the kinetic energy of the final nucleon. Left and right panels correspond to
the emission of protons and neutrons respectively. The solid histogram 
shows results without FSI and the dashed one the full model.}
\label{fig:charged}
\end{center}\end{figure}

The nucleons spectra produced by CC processes induced by muon neutrinos 
are shown in Fig.~\ref{fig:charged} for Argon. Of course neutrinos  
only interact via CC with neutrons and would emit protons, but these primary 
protons interact strongly with the medium and collide with other nucleons which
are also ejected. As a consequence there is  a reduction of the flux of high 
energy protons but a large number of secondary nucleons, many of them neutrons,
of lower energies appear.

\begin{figure}[htb]\begin{center}
\includegraphics[scale=0.3]{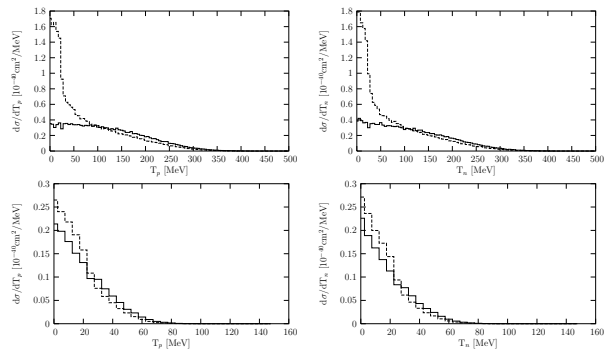}
\caption{\footnotesize Neutral current $^{40}O(\nu,\nu+N)$ at 500 MeV 
(upper panels) and 150 MeV (lower panels) cross sections as a function
of the kinetic energy of the final nucleon. Left and right panels correspond to
the emission of protons and neutrons respectively. The solid histogram 
shows results without FSI and the dashed one the full model.
}
\label{fig:neut}
\end{center}\end{figure}

The energy distributions of nucleons emitted after a NC interaction are shown
in Fig.~\ref{fig:neut}. Our results without rescattering compare well with 
other calculations like those of \cite{Martinez:2005xe}.
However, in these latter cases, which incorporate the nucleons final state
interaction via the use of optical potentials, the main effect is to reduce
the cross section at all energies instead of displacing the strength
towards lower energies as we find.  

\begin{figure}[htb]\begin{center}
\includegraphics[scale=0.3]{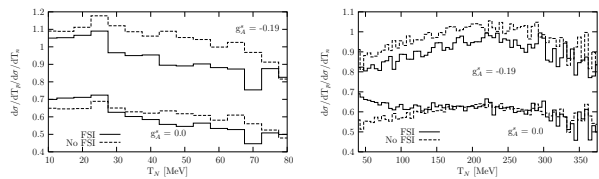}
\caption{\footnotesize Ratio of $d\sigma/dE$ for protons over that for
neutrons for $E_\nu=150 MeV$ and  $E_\nu=500 MeV$ in the reaction
$\nu+^{40}Ar\to \nu'+N+X$ as a function of the nucleon kinetic energy.}
\label{fig:rat}
\end{center}\end{figure}

The ratio of proton to neutron  quasielastic cross section could be very 
sensitive  to  the strange quark axial form factor of the nucleon, and thus
to the $g^s_A$ parameter. Our rescattering approach produces minor changes for
light nuclei, because of the smaller average density, and for low energies
because most secondary nucleons are below our 30MeV cut.
However, the sensitivity is larger for both heavier nuclei and for larger
energies of the neutrinos as shown in Fig.~\ref{fig:rat} 


\begin{thebibliography}{9}

\bibitem{Nie04} J. Nieves, J.E. Amaro and M. Valverde, 
                Phys. Rev. C 70 (2004) 055503.

\bibitem{LSND}  C. Athanassopoulos {\it et al.} , 
                Phys. Rev. C 55 (1997) 2078
                L.B. Auerbach {\it et al.} ,
                Phys. Rev. C 66 (1997) 0155501;
                
\bibitem{GNO97} A. Gil, J. Nieves and E. Oset, 
                Nucl. Phys. A 627 (1997) 543.

\bibitem{Al96}  W.M. Alberico, S.M. Bilenky, C. Giunti and C. Maieron,
                Z. Phys. C 70 (1996) 463.

\bibitem{Ga71}  S. Galster, {\it et al.} , 
                Nucl. Phys. B 32 (1971) 221.

\bibitem{Ga93}  G.T. Garvey, W.C. Louis and D.H. White, 
                Phys. Rev. C 48 (1993) 761.

\bibitem{COV94} R.C. Carrasco, M.J. Vicente Vacas and E. Oset,
                Nucl. Phys. A 570 (1994) 701. 

\bibitem{Martinez:2005xe}
                M.C. Martinez, P. Lava, N. Jachowicz, J. Ryckebusch, 
                K. Vantournhout and J.M. Udias,
                arXiv:nucl-th/0505008; 
                C.J. Horowitz, H.c. Kim, D.P. Murdock and S. Pollock,
                Phys. Rev. C 48 (1993) 3078;
                B.I.S. van der Ventel and J. Piekarewicz,
                Phys. Rev. C 69 (2004) 035501;
                A. Meucci, C. Giusti and F.D. Pacati,
                Nucl. Phys. A 744 (2004) 307.


\end{thebibliography}
\end{document}